\documentclass[3p]{elsarticle}

\usepackage{hyperref}

\journal{Sustainable Cities and Society, doi: \href{https://doi.org/10.1016/j.scs.2019.101602}{10.1016/j.scs.2019.101602}}

\usepackage{mathtools}
\usepackage{amssymb}
\usepackage[section]{placeins}
\usepackage{booktabs}
\usepackage[font=footnotesize,labelfont=bf,figurename=Fig.,labelsep=period,justification=justified,singlelinecheck=false]{caption}
\usepackage[caption=false]{subfig}
\usepackage{float}

\usepackage{etoolbox}
\makeatletter
\patchcmd{\ps@pprintTitle}
{Preprint submitted to}
{Accepted preprint,}
{}{}
\makeatother

\biboptions{authoryear}

\begin{document}

\begin{frontmatter}

\title{Streetscape augmentation using generative adversarial networks: insights related to health and wellbeing}

\author[addressUoM1]{Jasper S. Wijnands\corref{mycorrespondingauthor}}
\cortext[mycorrespondingauthor]{Corresponding author}
\ead{jasper.wijnands@unimelb.edu.au}

\author[addressUoM1]{Kerry A. Nice}
\author[addressUoM1]{Jason Thompson}
\author[addressUoM1]{Haifeng Zhao}
\author[addressUoM1,addressUoM2,addressUoM3]{Mark Stevenson}

\address[addressUoM1]{Transport, Health and Urban Design, Melbourne School of Design, The University of Melbourne, Parkville, Australia}
\address[addressUoM2]{Melbourne School of Engineering, The University of Melbourne, Parkville, Australia}
\address[addressUoM3]{Melbourne School of Population and Global Health, The University of Melbourne, Parkville, Australia}

\begin{abstract}
Deep learning using neural networks has provided advances in image style transfer, merging the content of one image (e.g., a photo) with the style of another (e.g., a painting). Our research shows this concept can be extended to analyse the design of streetscapes in relation to health and wellbeing outcomes. An Australian population health survey ($n$=34,000) was used to identify the spatial distribution of health and wellbeing outcomes, including general health and social capital. For each outcome, the most and least desirable locations formed two domains. Streetscape design was sampled using around 80,000 Google Street View images per domain. Generative adversarial networks translated these images from one domain to the other, preserving the main structure of the input image, but transforming the `style' from locations where self-reported health was bad to locations where it was good. These translations indicate that areas in Melbourne with good general health are characterised by sufficient green space and compactness of the urban environment, whilst streetscape imagery related to high social capital contained more and wider footpaths, fewer fences and more grass. Beyond identifying relationships, the method is a first step towards computer-generated design interventions that have the potential to improve population health and wellbeing.
\end{abstract}


\end{frontmatter}

\section{Introduction}
\label{sec:intro}

\subsection{Streetscape design}
\label{sec:ud}

Given the recent mass movement of people to cities, a trend expected to continue into the future \cite[][]{iom2015world}, neighbourhood planning is becoming increasingly important. Historically, as a response to problems related to overcrowding such as outbreaks of diseases following the industrial revolution, improving the health of city residents has been one of the main drivers of urban planning \cite[][]{sharifi2016from}. For example, in the early 20$^{\text{th}}$ century the Garden City movement \cite[][]{howard1902garden} aimed to address these issues by combining the best elements of city and countryside living in a single neighbourhood design. This was followed by the Neighbourhood Unit concept \cite[][]{perry1929neighborhood}, attempting to address social problems by developing self-contained neighbourhoods that increase pedestrian safety and provide a strong sense of community. However, according to \citet{mehaffy2015neighborhood}, the Neighbourhood Unit did not improve social interaction, health conditions or walking behaviour. Hence, although utopian visions aimed to improve health and wellbeing outcomes of city residents, not all were successful. Modernism concepts, such as Le Corbusier's Radiant City \cite[][]{corbusier1933ville,fishman1982urban}, introduced superblocks and high-rise buildings. In contrast, the Broadacre City \cite[][]{wright1932disappearing} reserved one-acre plots per family for living purposes. Single-use zoning had some advantages, such as ease of implementation. However, suburban development led to increased car dependency and, in reality, had adverse health effects including non-communicable disease. Later, neo-traditional movements focussed on walkability, public transport, compact form and medium-high density to address issues caused by suburbanisation \cite[][]{furuseth1997neotraditional,rutheiser1997beyond}. Eco-urbanism aimed for sustainable development, for example, using renewable energy technologies to reduce greenhouse gas emissions and protecting natural environments from new developments by revitalising existing urban areas instead \cite[][]{joss2013towards,tsolakis2015eco}. Several design concepts have been linked to sustainability, including compactness, sustainable transport, density, mixed land use, diversity, passive solar design and greening \cite[][]{jabareen2006sustainable}. The neo-traditional development, eco-city and compact city \cite[][]{dantzig1973compact} all incorporate some of these sustainable urban form concepts. These developments indicate that the historically strong link between urban form and health has been re-emerging.

Various recent studies have linked the urban form to health and wellbeing outcomes \cite[][]{devries2013streetscape, sallis2015is}. For example, hard urban surfaces and reduced shading can lead to heat island effects, with associated health risks such as heat exhaustion and heat stroke, especially impacting the elderly \cite[][]{smith2008designing, carmona2010public}. Urban green space has a cooling effect that could mitigate some of these issues \cite[][]{bowler2010urban}. The level and variability of streetscape greenness has further health impacts, e.g., on perceived general health \cite[][]{maas2006green} and the prevalence of cardiovascular disease \cite[][]{pereira2012association}. Survey-based approaches have identified that living in a green environment has positive associations with self-reported health, including the number of symptoms experienced in the last two weeks \cite[][]{devries2003natural}. Direct physical connections have been found as well, for example, \citet{wardthompson2012more} showed that a lack of green space in deprived communities is associated with increased stress levels as measured by salivary cortisol.

Urban form also influences the frequency of active transport such as walking and cycling \cite[][]{frank2001built,hensley2014healthy}, providing associated health benefits \cite[][]{turrell2013can}. A higher perceived confinement of space and sense of intimacy (as measured by visual enclosure) is related to increased pedestrian activity \cite[][]{yin2016measuring}. Specific streetscape features such as the proportion of windows on the street, the proportion of active street frontage and the amount of street furniture also significantly impact pedestrian activity \cite[][]{ewing2016streetscape}. Further, pavement condition is considered very important by pedestrians and poor conditions lead to strong dissatisfaction among elderly pedestrians, which can affect physical activity levels \cite[][]{stradling2007performance}.

The built environment also affects crime rates and perceived safety \cite[][]{foster2008built}. For example, \citet{salesses2013collaborative} found a significant association between violent crime and the perceptions of streetscapes related to safety and social class. Interventions that improve the appearance of abandoned buildings (e.g., by installing working doors and windows) and vacant lots (e.g., removing rubbish, planting grass and trees, installing low fences) can reduce the prevalence of firearm violence \cite[][]{branas2016urban}. The effect of greening vacant land also has mental health implications. Through a cluster randomised trial, \citet{south2018effect} found that greening significantly reduced feelings of depression and worthlessness of residents living nearby. \citet{cohen2008built} found that parks are positively associated with the perception of mutual trust and willingness of people to help each other. Further, the architectural features present in urban streetscapes have been found to affect wellbeing outcomes \cite[][]{spokane2007identifying}. A literature review by \citet{mair2008are} identified that depression was more consistently associated with measures of the built environment than with socioeconomic deprivation, residential stability or race composition.

The studies above indicate the perception of urban form has various health and wellbeing impacts. \citet{lynch1960image} investigated the human perception of urban form and concluded humans map their surroundings using elements such as paths, edges, districts, nodes and landmarks. Further, \citet{lozano1990community} defined urban form as the aggregation of more or less repetitive elements, such as block size and form, street design, street patterns and the layout of parks. Big data sources such as Google Street View (GSV) imagery \cite[][]{anguelov2010google} now allow for in-depth analyses of these urban features. For example, large numbers of geo-tagged photos have been used to detect patterns of urban usage and public perception of functional and social attributes \cite[][]{liu2016cimage, zhou2014recognizing}. \citet{doersch2012what} used geo-localised street level images to discover unique visual features in a city, such as balconies, windows with railings and special Parisian lampposts in Paris, France. Place Pulse, a database of urban imagery using crowd-sourced classifications, quantifies perceptions of urban areas, including safety, beauty, and liveliness \cite[][]{dubey2016deep, naik2014streetscore}. Various studies have also attempted to quantify the level of green space visible at the street level. For example, \citet{li2015assessing} used colour information in GSV images to estimate and map urban green space. Further, \citet{seiferling2017green} focussed on identifying tree cover from GSV images using computer vision techniques.

Although the research findings described throughout this introduction are important, the methods employed typically focus on a selection of urban factors or sometimes just one factor in isolation (e.g., greenness). As these factors are pre-supposed to be associated with outcomes, they are potentially subject to a degree of observer bias based on historical, theoretical, or disciplinary background. Further, when multiple factors are included, there is limited understanding of how these interact to improve health or how suggested changes to urban form might translate in the real world. For example, while traditional linear approaches such as regression models might include multiple observed factors that recommend a `30\% improvement in greenness', `25\% increase in density', and a `15\% increase in diversity', it is left to the imagination of residents, policy-makers and planners to understand how these changes might actually appear when translated into the urban landscape. Given the mixed success, historically, of utopian planning visions on city residents' health, the analysis of empirical streetscape data using an objective, unsupervised approach will identify elements of neighbourhood planning movements that can be linked to actual health improvements of residents.

\subsection{Style transfer}
\label{sec:style}

Compared to the traditional approaches described above, the current study does not require an a priori selection of features potentially subject to observer bias. Our approach is based on the concept of generative adversarial networks (GANs), first described in \citet{goodfellow2014generative}. GANs are based on game theory and consist of multiple competing models, namely generator and discriminator neural networks. The role of the generator is to generate images in the new style, while the discriminator assesses whether the generated image looks realistic. Using different formulations of the internal loss function and neural network structures, GANs have been adapted for various image translation tasks; for example, increasing image resolution to recover finer details (i.e., super-resolution) \cite[][]{ledig2017photo, sonderby2017amortised} or filling in missing regions of an image (i.e., semantic inpainting) \cite[][]{pathak2016context, yeh2017semantic}. Another application of image-to-image translation is style transfer, which entails merging the style of a collection of images and the content of another image. Examples of style transfer include translating (i) photos to paintings, (ii) black and white photos to colour, (iii) summer photos to winter and (iv) daytime photos to night \cite[e.g.,][]{gatys2016image, isola2017image, zhu2017unpaired}.

Importantly, these techniques can also be applied to images representing the design of the urban environment. As health and wellbeing outcomes have been shown to be related to the human perception of urban form (see Sect.~\ref{sec:ud}), the image style could be defined as streetscapes with either a positive or negative human perception. \citet{salesses2013collaborative} have shown the large variety in human perceptions of streetscapes captured in GSV images. By changing the style of the streetscape towards a more positive perception, it might be possible to also improve health or wellbeing outcomes. Note that not all variation in GSV images may be attributed to human perception or health and wellbeing outcomes. While for the style transfer domains described above (e.g., black and white to colour) the style is uniquely defined, the exact streetscape style that influences human perception and health outcomes still has to be discovered. Therefore, research is required that investigates if a GAN is able to successfully capture these styles.

Using GANs it may be possible to design by example rather than designing each urban space individually, based on large streetscape datasets. By sampling design of successful areas using outcomes such as liveability, non-communicable disease and mental health, unsupervised image-to-image translation may be able to apply key characteristics of these areas to any other location. This study provides an investigation along these lines, focussing on the differences in streetscape imagery from areas with good and bad perceived health and wellbeing outcomes. The proposed research has two objectives: (i) to present the application of a new technique for streetscape analysis based on big data and artificial intelligence; and (ii) to present findings of relationships between health outcomes and specific design elements discovered using this method.

\section{Methods}
\label{sec:method}

This study used GANs to augment the design of streetscapes captured in GSV images, as GSV provides an impression of streetscape design from an observer's point of view. The locations at which to obtain these GSV images were determined using survey-based health and wellbeing data.

\subsection{Health and wellbeing}

The data used to identify the health, wellbeing and demographic characteristics of areas of the Melbourne metropolitan region were derived from a Victorian Population Health Survey \cite[][]{department2017victorian}. This health survey is a population-weighted computer-aided telephone interview survey that collected information from 34,000 residents across the State of Victoria relating to self-reported health outcomes and health behaviours. For this research, health outcomes and behaviours relating to perceptions of general health, social capital and life satisfaction across metropolitan Melbourne were included. Specifically, individuals were asked to rate their perceptions of general health on a five-point Likert scale ranging from 1 (excellent) to 5 (poor). Social capital was measured by asking individuals to recall the number of people they had spoken to on the day before. Overall life satisfaction was estimated by asking individuals how satisfied they were with life overall, with ratings from 1 (very satisfied) to 4 (very dissatisfied). This was supplemented with population and housing density data for validation purposes.

\subsection{Google Street View}

A GSV dataset was created by sampling the Greater Melbourne area, Australia. Locations were determined using the nodes of vector lines in a street network dataset \cite[][]{psma2017psma} to exclude most indoor imagery. For each selected location, four GSV images were retrieved at headings of 0, 90, 180 and 270 degrees and the Google logo was removed for deep learning purposes. A sample of 100,000 images was inspected for consistency and the complete dataset was post-processed by removing remaining indoor images based on file size, resulting in 4.5 million GSV images at a 256$\times$256 pixels resolution. This dataset and the underlying methodology is explained in more detail in the linked \textit{Data in Brief} paper \cite[][]{nice2018melbourne}. 

For each outcome variable of interest, the 10\% most and least desirable locations were selected with respect to self-reported health outcomes (as defined above) and were grouped into two sets (i.e., domains). The resulting locations were used to select a corresponding image from the GSV dataset. If no GSV image was available within 50 meters of the requested location, the location was excluded, leading to approximately 160,000 images per outcome variable (i.e., 80,000 per domain). These images were used as input to the GAN.

\subsection{Generative adversarial networks}
\label{sec:gan}

Since no matched pairs of images exist, supervised learning is not feasible. However, GANs have proven to be successful for style transfer using unsupervised learning \cite[e.g.,][]{zhu2017unpaired}. Previous studies have considered `style' as generating black and white images in colour, or as a painting (see Sect.~\ref{sec:style}). In contrast, this study defines style as the health or wellbeing outcome so images can be translated, for example, from a bad to a good general health style (as defined in the population health survey). The main structure and layout of the GSV image can be considered the `content', such that any differences between the original and generated image could eventually be considered amendments that can be incorporated into the current urban environment (in future research).

\begin{figure}[ht]
	\includegraphics[trim = 19mm 166mm 192mm 232mm, clip, width=0.3\paperwidth]{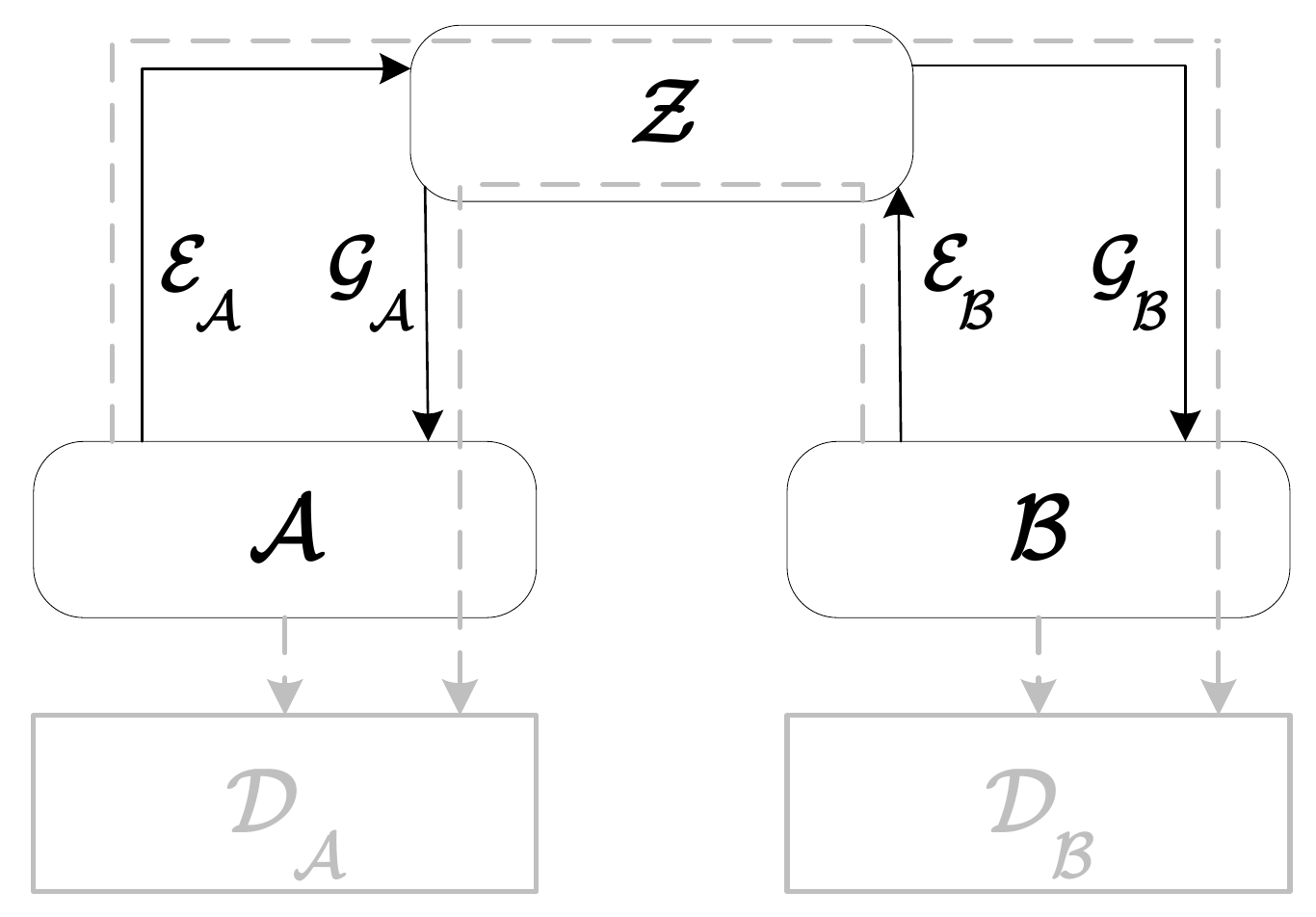}
	\caption{Illustration of model configuration.}
	\label{fig:modelConfig}
\end{figure}

Our approach is based on unsupervised image-to-image translation \cite[][]{liu2017unsupervised} and consists of two variational autoencoders (VAEs) \cite[][]{kingma2013auto} plus two discriminator models. The VAEs create a condensed representation of an input image in a latent space ($\mathcal{Z}$), which is assumed to be shared between the two domains ($\mathcal{A}$ and $\mathcal{B}$). The model configuration is illustrated in Fig.~\ref{fig:modelConfig}. Encoding functions $\mathcal{E}_\mathcal{A}$ and $\mathcal{E}_\mathcal{B}$ extract features from images of domain $\mathcal{A}$ and $\mathcal{B}$, respectively. Generator functions $\mathcal{G}_\mathcal{A}$ and $\mathcal{G}_\mathcal{B}$ decode the condensed representation from $\mathcal{Z}$ to the respective domain. Hence, the content is the information stored in $\mathcal{Z}$, while the style is embedded in generator functions $\mathcal{G}_\mathcal{A}$ and $\mathcal{G}_\mathcal{B}$. For example, elements that occur in GSV images of both domains (e.g., people) will likely be stored in $\mathcal{Z}$ without large differences between $\mathcal{G}_\mathcal{A}$ and $\mathcal{G}_\mathcal{B}$ (although this is not directly enforced), while elements unique to one of the image sets (e.g., green space) will only be captured in the generator function for that specific domain.

In this model configuration, images can be encoded in the shared latent space and decoded in either the original domain (i.e., a perfect model will reconstruct the same input image) or the alternative domain. Ideally, images would also be accurately regenerated when processed from $\mathcal{A}$ to $\mathcal{Z}$, to $\mathcal{B}$, to $\mathcal{Z}$, to $\mathcal{A}$ and, similarly, $\mathcal{B} \rightarrow \mathcal{Z} \rightarrow \mathcal{A} \rightarrow \mathcal{Z} \rightarrow \mathcal{B}$ (cycle consistency). The discriminator models ($\mathcal{D}_\mathcal{A}$ and $\mathcal{D}_\mathcal{B}$) assess the quality of images translated to the alternative domains. The associated loss function combines all discriminator, reconstruction and cycle consistency losses using hyperparameters $\lambda_0$, ($\lambda_1$, $\lambda_2$) and ($\lambda_3$, $\lambda_4$), respectively \cite[see][]{liu2017unsupervised}. These hyperparameters are used to control which of the loss components are emphasised during model training.

Various experiments with different hyperparameter values were performed to achieve a balanced translation. For example, when prioritising generator losses (either reconstruction or cycle consistency) translated images were very similar to the original, which does not allow for the identification of relationships between streetscape design and health outcomes. In contrast, when style transfer is too strong, the resulting image may appear abstract or the changes cannot be considered `amendments' to the urban environment, but a complete redesign. Compared to the default values, the selected hyperparameters (($\lambda_0$, $\lambda_1$, $\lambda_2$, $\lambda_3$, $\lambda_4$) = (50, 0.1, 100, 0.1, 100)) place more emphasis on the discriminator loss, prioritising the generation of more realistic images in the style domain over perfect reconstruction and cycle consistency.

Pre-processing steps included the reduction of image resolution to 128$\times$128 pixels to speed up computation. The use of higher-resolution images with GANs is possible \cite[][]{curto2017high}, but was outside the scope of this study. GANs were then calibrated using backpropagation for 1.2 million iterations (approximately 60 hours of training on an NVIDIA GTX 1080 GPU), with alternating weight updates to the generator and discriminator networks.

\subsection{Analysis of generated images}

The calibrated model was used to translate images from one domain to the other. Generated images were analysed in the following three ways: (i) a visual comparison using difference images; (ii) using statistical measures; and (iii) the computation of the average translation, providing insights into the neural network's inference patterns.

\paragraph{Visual comparison}

To compare a generated image to its original, pixel-based difference images were constructed using ImageMagick \cite[][]{imagemagick2018imagemagick}. The visual analysis of a large sample of translated images identified several themes per outcome variable. In Sect.~\ref{sec:res}, eight translated images are presented for most of the outcome variables, which are representative of these themes.

\paragraph{Statistical measures}

Statistical analysis was used to estimate the strength of the association between investigated outcome variables and local urban form. Several hundred images for each combination of outcome (i.e., general health, social capital, life satisfaction) and domain (i.e., low or high) were randomly selected. Each image was translated to the opposite domain and a difference image was constructed. Since the GAN first creates a high-level representation of an image in the shared latent space, details such as the exact colour value and positioning are lost. Therefore, even if part of the image is unaffected by the new style, almost all pixel values of the image change during translation. In order to identify the extent to which an image was changed, small changes were discarded (ImageMagick \textit{convert} function with parameters compose:difference, fuzz:5\%, transparent:black). The average proportion of non-white pixels gives an indication of the amount of change observed using the calibrated model.

The difference between the original and translated images was also investigated using standard statistical approaches; specifically, the mean squared error (MSE), peak signal-to-noise ratio (PSNR) \cite[e.g.,][]{bing2015assessing} and structural similarity index (SSIM) \cite[][]{wang2004image}. These statistics compute the average similarity between the original and translated image. MSE and PSNR focus on pixel-based differences, while SSIM is geared towards the degradation of structural information. Specifically, SSIM combines structure, luminance and contrast measures for many small, corresponding patches in both images. The GAN translates the style of the GSV image and attempts to preserve its content at the same time. Therefore, it is expected that perceived changes in structural information will be limited by construction, while pixel-based differences could be more indicative of the amount of style transfer. For example, translation of a road surface from gravel to sealed road will lead to large pixel-based differences captured by MSE and PSNR, while the image structure could still be quite similar.

\paragraph{Average translation}

Besides quantifying the amount of change occurring during the translation process, it is also possible to get a better understanding of the type of changes the GAN makes. Therefore, the average translation by the generator neural network was computed using all domain images. This can provide evidence on whether the identified themes hold overall. First, the average GSV image was computed pixel-by-pixel for corresponding pixels of all 80,000 images in a domain. Then, all these images were translated using the calibrated GAN and the average translated image was computed. Differences between the average original and translated image give an insight into what the network prioritises when converting one style to another and exemplifies the inner workings of the generator neural network.

These three different methods are employed as the validity of results may be difficult to assess, since the theoretical foundation of associations between urban form and health and wellbeing outcomes is not captured by the proposed methodology (i.e., the method is based on large-scale empirical data). In addition, the face validity (e.g., consistency with expected performance) of the model is first investigated using domains that are readily understood: (i) density; and (ii) city/park translations.

\section{Results}
\label{sec:res}

\subsection{Visual comparison -- density, city/park}
\label{sec:construct_val}

\paragraph{Density translation}

\begin{figure}[H]
	\makebox[\textwidth][c]{
		\subfloat[Density]{
			\fbox{\includegraphics[trim = 0mm 0mm 0mm 0mm, clip, width=0.39\paperwidth]{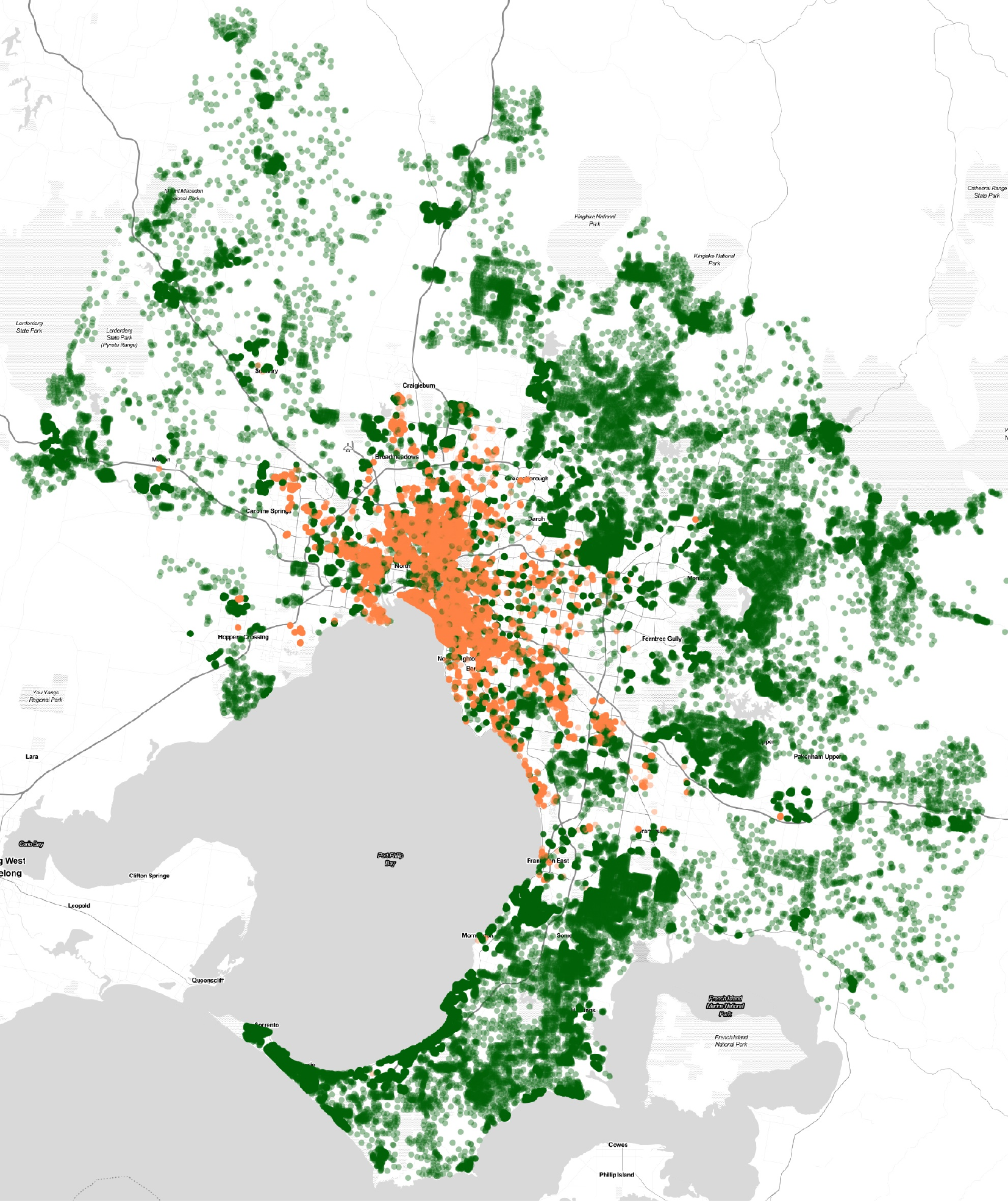}}
			\label{fig:domains_1a}
		}
		~
		\subfloat[City and park]{
			\fbox{\includegraphics[trim = 0mm 0mm 0mm 0mm, clip, width=0.39\paperwidth]{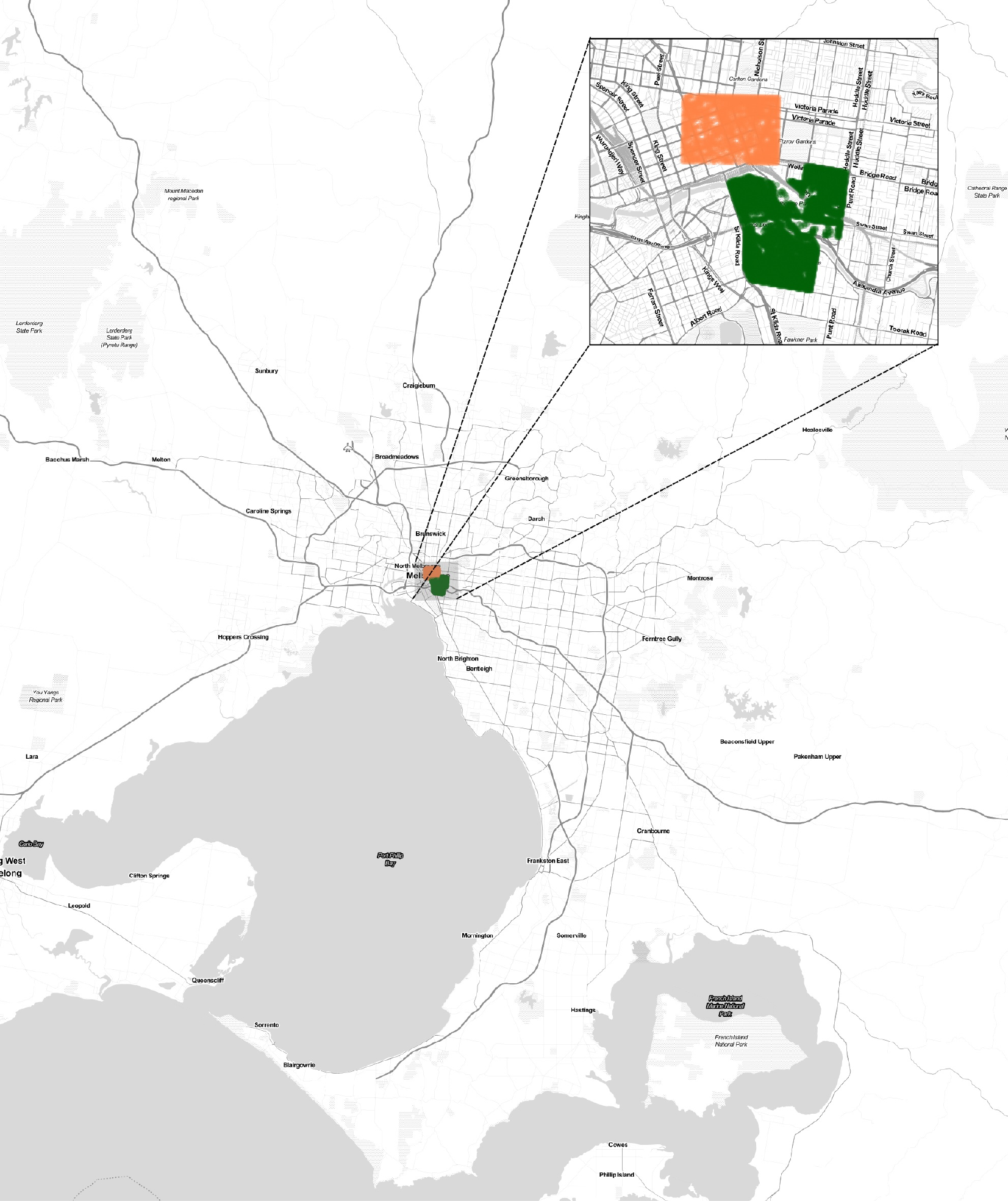}}
			\label{fig:domains_1b}
		}
	}
	\caption{Selected domains: (a) locations with the highest (red/light grey) and lowest (green/dark grey) 10\% of density, and (b) city (red/light grey) and park (green/dark grey) locations. Figures were generated using QGIS \cite[][]{qgis2018}.}
	\label{fig:domains_1}
\end{figure}

\begin{figure}[ht]
	\makebox[\textwidth][c]{\includegraphics[trim = 0mm 0mm 0mm 0mm, clip, width=0.9\paperwidth]{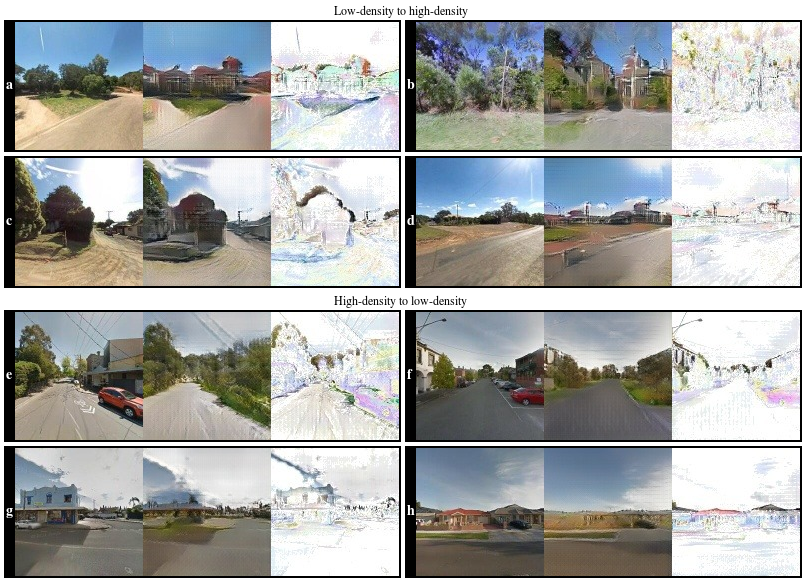}}
	\caption{Sample images showing translation of low-density GSV images to a high-density style (sub-figures a--d) and vice versa (e--h). Each sub-figure shows the original GSV image (left; source of map data: Google), the generated image (middle) and their pixel-by-pixel differences (right).}
	\label{fig:val_dens}
\end{figure}

Fig.~\ref{fig:domains_1}a shows the selected locations corresponding to the highest and lowest density in the Greater Melbourne area, with the highest density observed near Melbourne's central business district (CBD). After model training, the GAN reconstructs images accurately when encoding in the latent space and regenerating (results not presented). Fig.~\ref{fig:val_dens} shows the original GSV and translated images from low to high-density and vice versa. Specifically, the translation from low to high-density showed the following themes: (i) the creation of new buildings, generally in areas currently occupied by trees; and (ii) the conversion of grass and gravel surfaces to concrete (see Figs.~\ref{fig:val_dens}a--d). The reverse translation of high to low-density can be characterised by (i) the removal of buildings, cars, street lights, road markings and road signs, which are generally replaced by grass, trees or bushes; (ii) the conversion of multi-level buildings to single-level; and (iii) the creation of larger open spaces or plains (see Figs.~\ref{fig:val_dens}e--h).

\paragraph{Park and city}

\begin{figure}[ht]
	\makebox[\textwidth][c]{\includegraphics[trim = 0mm 0mm 0mm 0mm, clip, width=0.9\paperwidth]{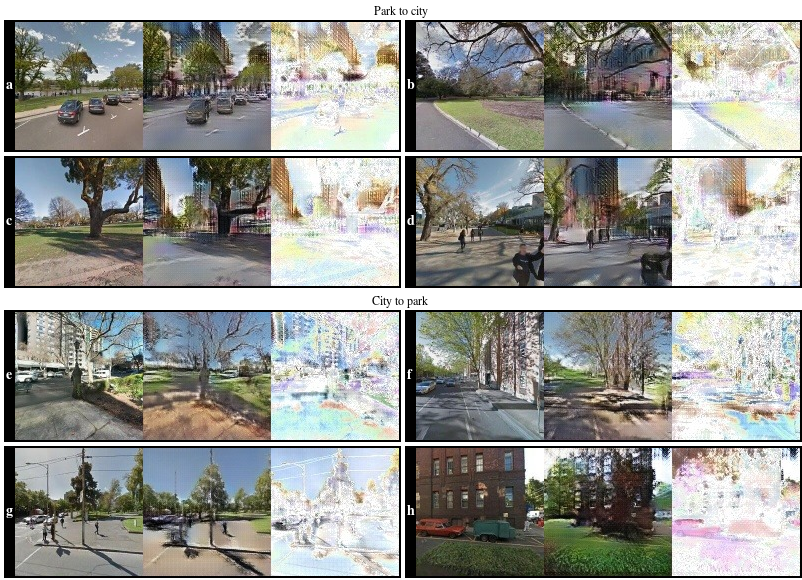}}
	\caption{As Fig.~\ref{fig:val_dens}, but for the translation of park GSV images to a city style (sub-figures a--d) and vice versa (e--h).}
	\label{fig:val_park_city}
\end{figure}

A more extreme transformation is the translation between Melbourne's inner-city and parks (i.e., Kings Domain and the Royal Botanic Gardens), as pictured in Fig.~\ref{fig:domains_1}b. These domains have fewer images than the survey-based domains, namely about 6200 city and 13,800 park GSV images. Various themes were observed for the park to city translation, including (i) the reduction of open spaces by creating high-rise buildings; and (ii) the replacement of natural surfaces such as grass and gravel by concrete (see Figs.~\ref{fig:val_park_city}a--d). Other features, such as cars, trees and pedestrians, are maintained. The reverse translation also shows several themes, namely (i) the generation of additional green space; (ii) the conversion of sealed surfaces to gravel; and (iii) the elimination of vehicles, buildings and power lines, which are scarcely observed in park GSV images (see Figs.~\ref{fig:val_park_city}e--h).

Fig.~\ref{fig:val_park_city}h shows the translation of a large facade to multiple trees, although the translated image still contains remnants of the original building. Importantly, this indicates that large facades do not (or, rarely) occur in the park domain, leading to complete replacement.

The changes described above are in line with expectations when translating between low and high-density, and city and park domains, respectively. Translating from low to high-density leading to the addition of buildings is self-evident, as is building removal for the opposite translation. Similarly, the creation of natural surfaces in a park style and the addition of high-rise buildings in an inner-city style is clear. Visually the results have high face validity, providing confidence in the proposed method. Specifically, these experiments highlight that GANs can capture differences between selected image sets as distinct styles, even though streetscapes are heterogeneous and the image sets contain significant overlap (e.g., low-density areas may have some multi-level buildings) and not every element in the GSV image may be related to concepts as `city' and `park'. These results allow us to move forward and explore the transformation of health-related domains, where the captured styles are less obvious.

\FloatBarrier

\subsection{Visual comparison -- health and wellbeing}
\label{sec:theme}

\begin{figure}[H]
	\makebox[\textwidth][c]{
		\subfloat[General health]{
			\fbox{\includegraphics[trim = 0mm 0mm 0mm 0mm, clip, width=0.39\paperwidth]{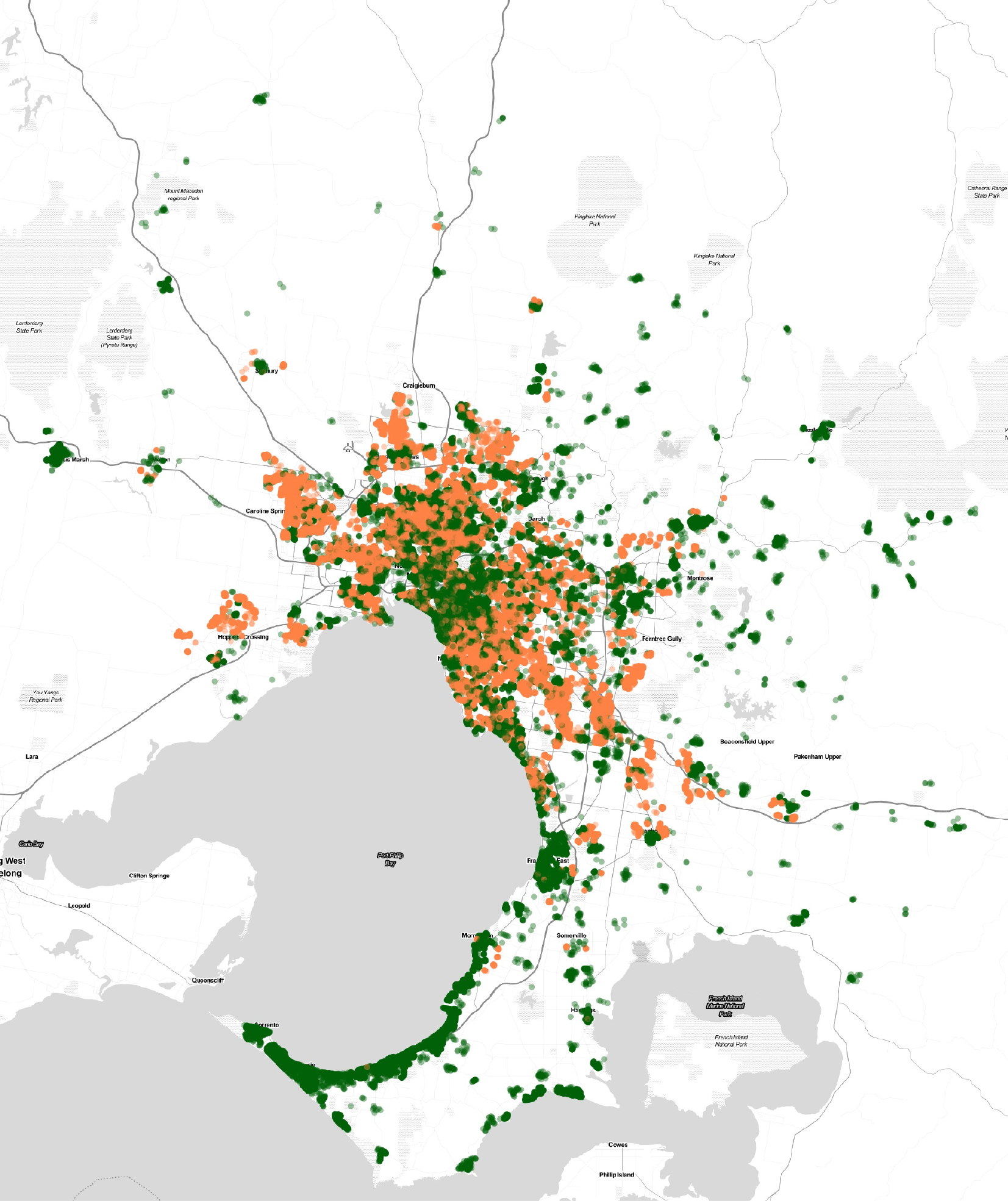}}
			\label{fig:domains_2a}
		}
		~
		\subfloat[Social capital]{
			\fbox{\includegraphics[trim = 0mm 0mm 0mm 0mm, clip, width=0.39\paperwidth]{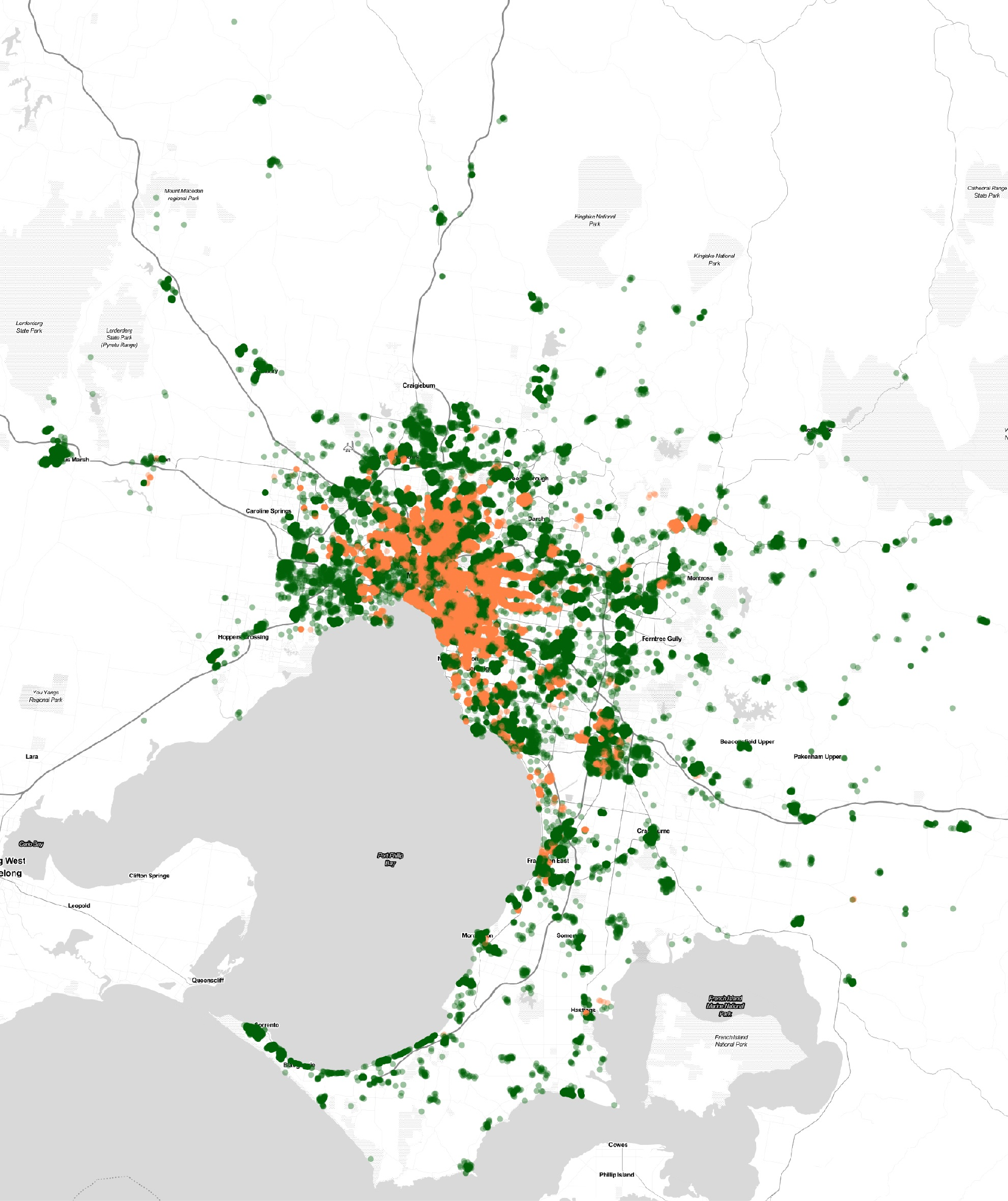}}
			\label{fig:domains_2b}
		}
	}
	
	\makebox[\textwidth][c]{
		\subfloat[Life satisfaction]{
			\fbox{\includegraphics[trim = 0mm 0mm 0mm 0mm, clip, width=0.39\paperwidth]{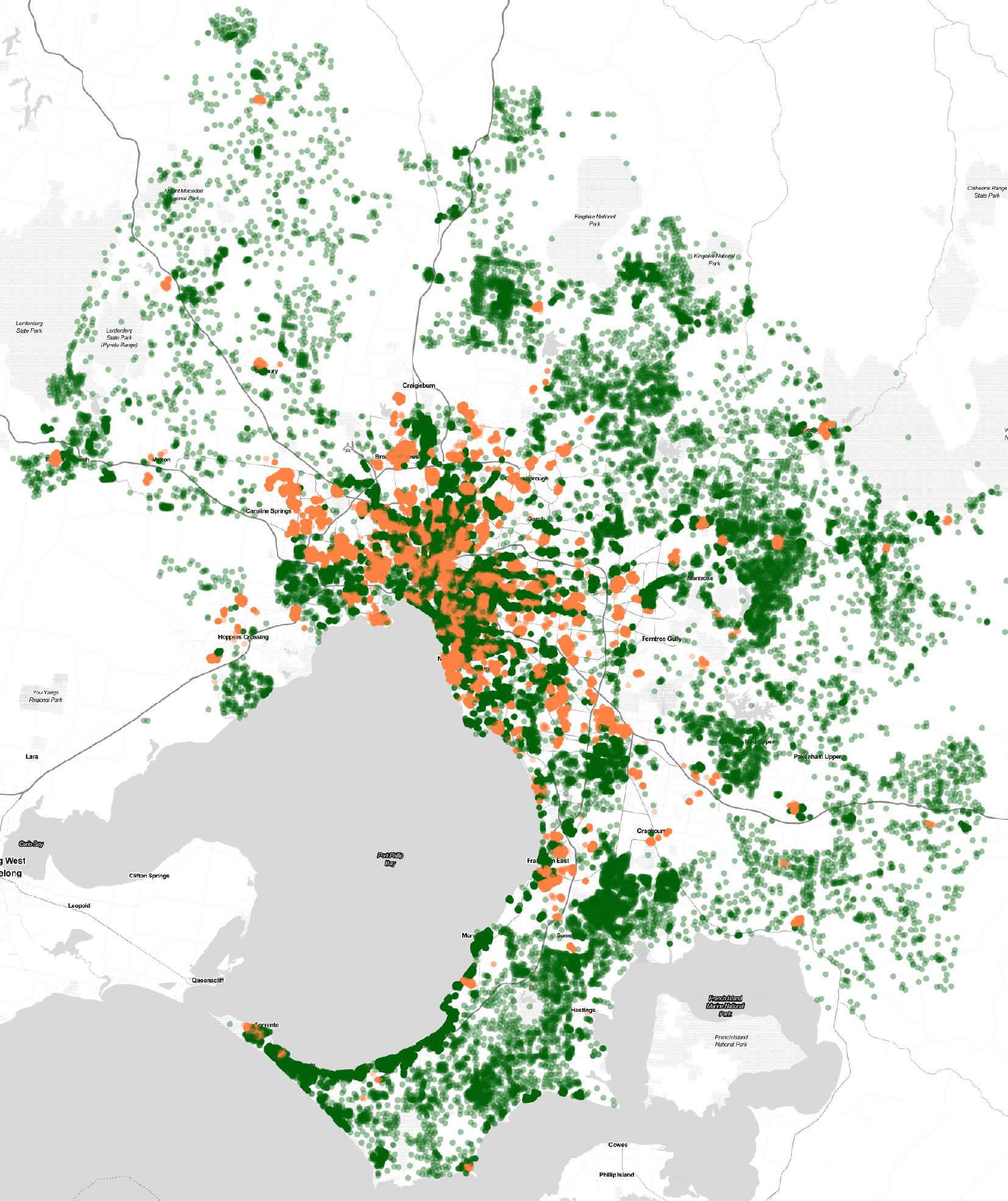}}
			\label{fig:domains_2c}
		}
		~
	}
	\caption{Selected domains: locations with the best 10\% (green/dark grey) and worst 10\% (red/light grey) of (a) general health, (b) social capital, and (c) life satisfaction.}
	\label{fig:domains_2}
\end{figure}

The selected locations for general health, social capital and life satisfaction are presented in Fig.~\ref{fig:domains_2}. Some clustering of the domains is apparent, especially for general health and social capital. Self-reported general health is better in the CBD and coastal suburbs, while social capital is higher in the suburbs. For life satisfaction, the locations corresponding to the best responses are distributed over the Greater Melbourne area.

\paragraph{General health}

\begin{figure}[ht]
	\makebox[\textwidth][c]{\includegraphics[trim = 0mm 0mm 0mm 0mm, clip, width=0.9\paperwidth]{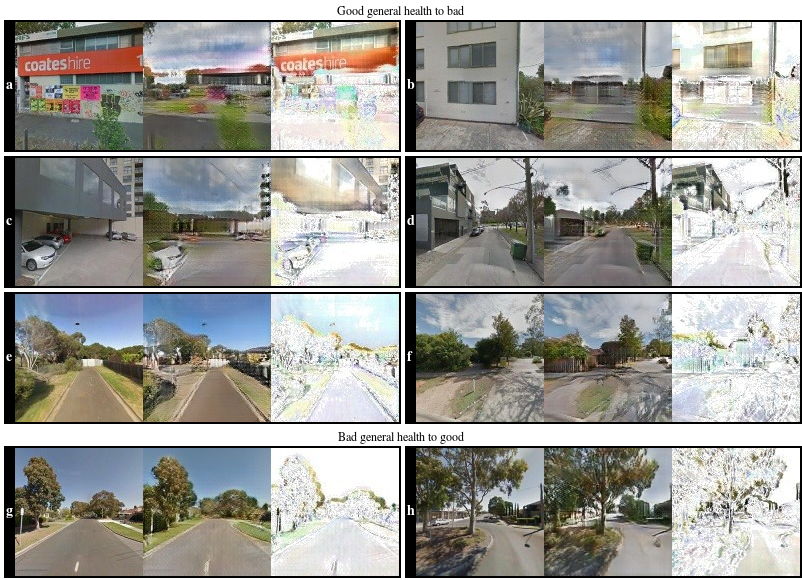}}
	\caption{As Fig.~\ref{fig:val_dens}, but for the translation of GSV images at locations with good general health to a style related to bad general health (sub-figures a--f) and vice versa (g--h).}
	\label{fig:gen_health}
\end{figure}

The translation from a good to a bad general health style shows that buildings appearing close to the GSV camera are moved further away (see Figs.~\ref{fig:gen_health}a--d). This indicates that compactness of the urban environment in Melbourne is mostly observed in areas with good general health. GSV images with nearby structures are generally not present in the bad general health domain, causing buildings to be moved and sky to appear during the translation. Figs.~\ref{fig:gen_health}e--f exemplify the reduction in green space for the same translation, where trees become smaller and grass areas are reduced or removed. Natural surfaces, such as gravel footpaths in Fig.~\ref{fig:gen_health}f, become sealed. The reverse translation (Figs.~\ref{fig:gen_health}g--h) supports these findings, showing more grass and larger, denser trees.

\paragraph{Social capital}

\begin{figure}[ht]
	\makebox[\textwidth][c]{\includegraphics[trim = 0mm 0mm 0mm 0mm, clip, width=0.9\paperwidth]{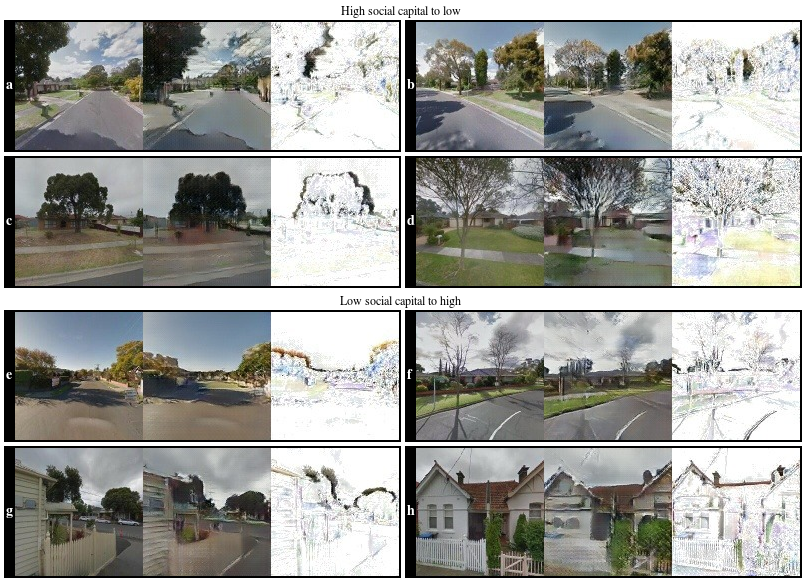}}
	\caption{As Fig.~\ref{fig:val_dens}, but for the translation from high to low social capital (sub-figures a--d) and vice versa (e--h).}
	\label{fig:talk}
\end{figure}

The translations related to social capital show that grass areas disappear, while trees become taller and more dense, when translating to the style based on low social capital GSV images (see Figs.~\ref{fig:talk}a--d). The opposite pattern is apparent in the reverse translation, as exemplified in Fig.~\ref{fig:talk}e. Interestingly, the identified themes for general health show a change in the amount of green space (i.e., grass, bushes and trees), while this is not the case here: \textit{more} grass and \textit{smaller} trees appear in GSV images related to high social capital. Wider and/or new footpaths (see Figs.~\ref{fig:talk}e--f) and the removal of fences (see Figs.~\ref{fig:talk}g--h) indicate that the selected image sets also differ for these features.

\paragraph{Life satisfaction}

For life satisfaction no clear themes were observed, except a somewhat similar green space pattern as for social capital.

\newpage
\vspace*{-3.4cm}

\FloatBarrier

\subsection{Statistical measures}

Table~\ref{tab:change} shows the amount of change (i.e., the proportion of non-white pixels in difference images) observed following translation. Further, Table~\ref{tab:stats} investigates the difference between the original and translated images using MSE, PSNR and SSIM. Note that an image-to-image translation showing many changes corresponds to a high MSE, low PSNR and low SSIM. Both Tables~\ref{tab:change} and \ref{tab:stats} show that the largest changes occur for the park to city conversion, with on average the highest proportion of changed pixels, highest MSE and lowest PSNR. These domains contain minimal overlap (i.e., most images from the city domain are distinctly different from the images in the park domain), which results in larger changes when translating an image from park to city and vice versa. Other domains, especially life satisfaction, show more overlap (also see the lower degree of clustering in Fig.~\ref{fig:domains_2c}). For life satisfaction, the translated images are most similar to the original images with, on average, the lowest proportion of changed pixels, lowest MSE and highest PSNR. This is in line with the lack of themes identified for this variable during the visual analysis of the generated images (see Sect.~\ref{sec:theme}). Therefore, it seems that life satisfaction is not as strongly related to local urban form as the other investigated outcome variables.

\begin{table}[h]
  \caption{Average proportion of pixels that changes during translation.}
  \label{tab:change}
  \begin{tabular}{l l l}
    \toprule
    & Low to high & High to low\\
    \midrule
    Density & 53.2\% & 57.8\%\\
    Park / city & 86.1\% & 80.9\%\\
    General health & 44.2\% & 54.2\%\\
    Social capital & 42.2\% & 39.5\%\\
    Life satisfaction & 35.5\% & 37.3\%\\
    \bottomrule
  \end{tabular}
\end{table}

\begin{table}[h]
	\caption{Average MSE, PSNR (dB) and SSIM statistics between original and translated images. A higher MSE, lower PSNR and lower SSIM indicate less similarity.}
	\label{tab:stats}
			\begin{tabular}{l r r r r r r r r r}
				\toprule
				& \multicolumn{3}{c}{Low to high} & \multicolumn{3}{c}{High to low} & \multicolumn{3}{c}{Average}\\
				\cmidrule(r){2-4} \cmidrule(r){5-7} \cmidrule(r){8-10}
				& MSE & PSNR & SSIM & MSE & PSNR & SSIM & MSE & PSNR & SSIM\\
				\midrule
				Density & 942 & 19.0 & 0.57 & 1151 & 18.0 & 0.55 & 1047 & 18.5 & 0.56\\
				Park / city & 2191 & 15.1 & 0.59 & 1149 & 17.9 & 0.65 & 1670 & 16.5 & 0.62\\
				General health & 811 & 19.6 & 0.63 & 1055 & 18.5 & 0.61 & 933 & 19.0 & 0.62\\
				Social capital & 893 & 19.0 & 0.58 & 876 & 19.2 & 0.60 & 885 & 19.1 & 0.59\\
				Life satisfaction & 695 & 20.1 & 0.65 & 673 & 20.4 & 0.64 & 684 & 20.3 & 0.64\\
				\bottomrule
			\end{tabular}
\end{table}

\FloatBarrier

\subsection{Average translation}

\begin{figure}[ht]
	\makebox[\textwidth][c]{\includegraphics[trim = 0mm 0mm 0mm 0mm, clip, width=0.9\paperwidth]{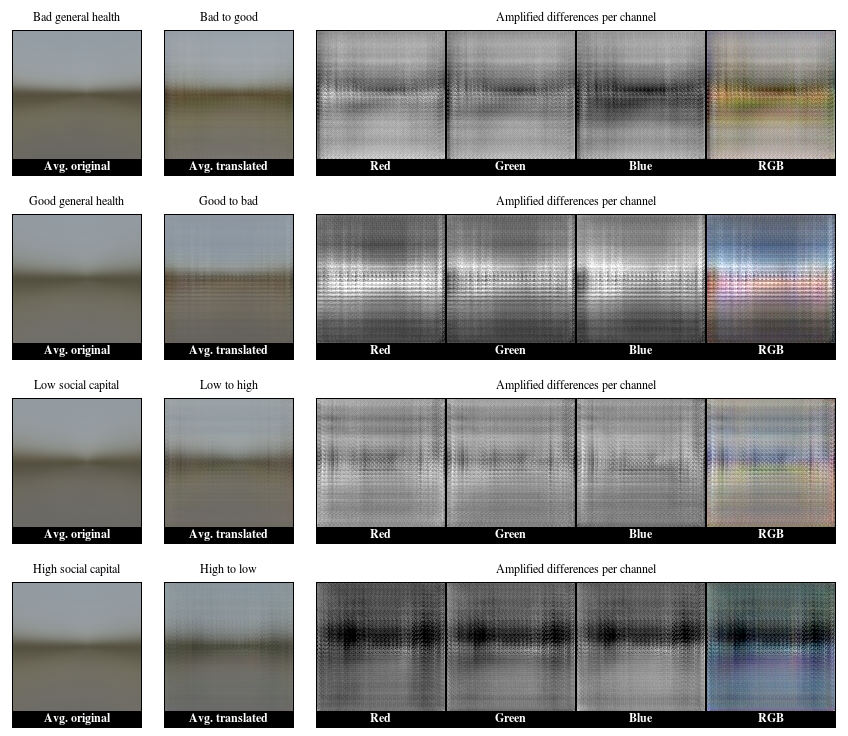}}
	\caption{Average changes the generator neural network makes for different style conversions. For the red, green and blue channel differences, white indicates an increase in that channel, grey (128, 128, 128) no change and black a reduction.}
	\label{fig:changes}
\end{figure}

Fig.~\ref{fig:changes} shows the average translations for general health and social capital. The changes in red, green and blue channels have been amplified to better visualise the spatial spread of changes in each channel. The RGB differences for the translation of images from bad general health locations to a good health style show more natural colours (i.e., yellow, green) below the horizon, supporting the findings in the visual analysis on the generation of additional grass and natural surfaces. The opposite translation shows more blue above the horizon, consistent with the visual analysis where buildings close to the GSV camera were partly replaced by sky and moved further away. The average translation to a high social capital style indicates changes mainly occur near the horizon, with more grass and more sky because of changes in vegetation (in line with the visual analysis). Finally, the reverse translation was linked to the replacement of grass areas by concrete, which is also exemplified by the blue colours below the horizon in Fig.~\ref{fig:changes}.

\FloatBarrier

Similar to the analysis above, the average pixel value and change in red, green and blue channels were calculated for all outcome variables and translations (see Table~\ref{tab:colour}). The statistics show an increase in the green (and red) channels for translations towards styles related to low density, parks, good general health, high social capital and high life satisfaction. This confirms various findings of the visual analyses regarding increased green space (see Sect.~\ref{sec:construct_val} and \ref{sec:theme}).

\begin{table}[h]
	\caption{Changes in colour channels for the average image per domain.}
	\label{tab:colour}
			\begin{tabular}{l c c c c c c c c c}
				\toprule
				& \multicolumn{3}{c}{Original} & \multicolumn{3}{c}{Translated} & \multicolumn{3}{c}{Difference (\%)}\\
				\cmidrule(r){2-4} \cmidrule(r){5-7} \cmidrule(r){8-10}
				& Red & Green & Blue & Red & Green & Blue & Red & Green & Blue\\
				\midrule
                Density - low to high & 125 & 127 & 121 & 123 & 124 & 118 & -1.7 & -2.5 & -2.1\\
                Density - high to low & 123 & 123 & 118 & 128 & 127 & 118 & 4.6 & 3.4 & -0.6\\
				\midrule
                Park to city & 114 & 116 & 108 & 94 & 93 & 91 & -17.8 & -19.5 & -16.3\\
                City to park & 107 & 108 & 104 & 109 & 111 & 106 & 1.8 & 2.9 & 1.7\\
				\midrule
                General health - low to high & 124 & 125 & 120 & 128 & 128 & 119 & 3.5 & 2.2 & -0.4\\
                General health - high to low & 123 & 124 & 119 & 123 & 124 & 121 & -0.1 & -0.3 & 1.5\\
				\midrule
                Social capital - low to high & 121 & 122 & 117 & 124 & 125 & 120 & 3.2 & 2.6 & 2.3\\
                Social capital - high to low & 124 & 125 & 120 & 114 & 118 & 116 & -7.3 & -5.1 & -3.4\\
				\midrule
                Life satisfaction - low to high & 124 & 125 & 120 & 128 & 130 & 122 & 3.3 & 3.8 & 1.8\\
                Life satisfaction - high to low & 122 & 123 & 118 & 117 & 122 & 115 & -4.0 & -0.7 & -3.1\\
				\bottomrule
			\end{tabular}
\end{table}

\FloatBarrier

\section{Discussion and conclusion}
\label{sec:disc}

This study presented an innovative approach to understanding variations in streetscape design through computer-generated visualisations. As described in the introduction (see Sect.~\ref{sec:ud}), the ideal urban form has been a point of discussion for more than a century. Utopian visions have been formulated as a response to issues of that time, leading to new developments or revitalisation of current urban areas. The impacts of these new conceptual frameworks are difficult to evaluate and are generally only visible after many years. The results presented in this paper are not derived from theory-driven urban planning concepts, but from a completely different perspective: the empirical exploration of unique features in large streetscape image datasets using artificial intelligence. When translating images to a new `style', the method automatically focusses on features that differentiate the image sets of selected domains (i.e., unsupervised learning). These translations show that areas in Melbourne with good general health are characterised by sufficient green space, compactness of the urban environment and natural surfaces. Further, streetscape imagery representative of neighbourhoods where residents had high levels of social capital contained more and wider footpaths, fewer fences and more grass. The themes observed in the translated imagery, such as compactness and greening, are pointing towards sustainable design philosophies as the compact city and eco-city as viable ways to enhance the health of city residents.

To identify if GANs are capable of capturing health and wellbeing styles, the analysis was expanded by including statistical measures and the average translation, supporting the conclusions. Findings are also in line with previous research, providing confidence in the new method and the validity of its results. For example, the positive impact of green space on general health was also identified by various other studies \cite[e.g.,][]{groenewegen2006vitamin, maas2006green, tzoulas2007promoting}. In addition, the positive health impact of compactness of the urban environment was quantified by \citet{stevenson2016land}, finding compact city design yields health gains of 420--826 disability-adjusted life-years per 100,000 population. With respect to non-compact urban form, various studies indicate urban sprawl has negative effects on public health, caused by increases in air pollution, transport-related crashes and reduced physical activity \cite[e.g.,][]{frumkin2002urban, gilescorti2012increasing}. In contrast, the relationship between urban sprawl and social capital is less clear. Sprawl could support some types of social capital while negatively impacting others \cite[][]{nguyen2010evidence}, indicating why compactness of the city was not observed in the social capital analyses. More green and other outdoor spaces, on the other hand, are associated with higher social capital. For example, \citet{sullivan2004fruit} found that suitable outdoor green space leads to a significant increase in the amount of social activity that takes place in these areas. Further, community gardens lead to increased social cohesion and social support \cite[][]{kingsley2006dig}. Perceived greenness of the urban environment is correlated with recreational walking, social cohesion and local social interaction \cite[][]{sugiyama2008associations}. Further, \citet{leyden2003social} found that walkable neighbourhoods lead to higher social capital. This is consistent with our findings relating social capital to outdoor space and infrastructure (i.e., grass and footpaths), although findings related to trees could be indicative of more remote areas.

This indicates a limitation of this study, as our method (as many other studies) does not provide causality of the identified relationships. For example, for both general health and social capital changes in green space were found, while \citet{maas2009social} indicate that loneliness and perceived shortage of social support could be a mediator for the relation between green space and health. Outcomes such as general health are influenced by many factors; the design of streetscapes is not their only driver. For example, the urban form at a larger scale, including land use mix and access networks, is not captured in a streetscape image. In addition, many factors not related to design (e.g., genetics, education, diet) influence these outcomes. However, if there would be no relationship between the design of streetscapes and health and wellbeing outcomes, changes between the original and generated images would be minimal without any clear, outcome-dependent themes. As shown in this research, this is not the case. In addition to imagery, future studies could aim to explicitly incorporate these other factors (e.g., distance to public transport, amenities, neighbourhood walkability) in the modelling. Various studies stress the need to change policies and amend local urban environments as a method to improve health and safety, as the impact of local environment improvements may be larger than interventions targeted at individuals \cite[e.g.,][]{kondo2015nature}. Eventually, GAN-based methods could be capable of suggesting amendments to specific streetscapes that improve health and wellbeing outcomes, by generating targeted interventions that also retain the current infrastructure of the local environment.

Note that images in this study were obtained for Melbourne, Australia and applicability of the findings to other international cities has not been investigated. Validity of these findings beyond Melbourne could be investigated using a similar approach, but different input imagery and health outcome data. A further limitation is that not all images are of great quality and some contain artefacts; for example, the reconstruction of cars could be more accurate. Inaccuracies in these renderings do not influence the major themes identified through visual analysis in this research.

We identify several additional avenues for future research. First, GAN architectures based on multimodal unsupervised image-to-image translation \cite[][]{huang2018multimodal, zhu2017toward} can generate multiple realisations based on a single input image, as there are generally multiple valid solutions when translating to a different domain. For the theme analysis performed in our study, multimodal translation was not required, but it could be explored in future research. Second, the Place Pulse 2.0 database \cite[][]{dubey2016deep} captured the perception of GSV images using crowd-sourcing. This directly measures the perception of a specific urban environment and could replace the survey part of our method. However, note that the Place Pulse study only measures a few factors (e.g., safety, beauty), limiting the options for streetscape analysis, while the style definition in our study is flexible and not necessarily restricted to health and wellbeing outcomes. Finally, the new method could eventually be used for virtual reality applications, where computer-generated design allows for an immersive and cost-effective evaluation of citywide design interventions, without the need for a potentially slower, more expensive, bottom-up design approach. Note that consecutive images along a single street were rendered in a consistent manner (results not presented).

Overall, this study has utility for researchers, urban planners and urban designers seeking to understand the perception and impact of streetscape design. It may, in the future, directly assist designers and policy-makers to implement our understanding of these relationships using targeted design interventions to improve the health and wellbeing of city residents. Finally, our study provides a tool that could be used in local urban design initiatives to inspire citizen-stakeholders who take an active role in designing their neighbourhood.

\section*{Acknowledgements}
JT is supported by an Australian Research Council Discovery Early Career Research Award [grant number DE180101411]. MS is supported by a National Health and Medical Research Council (Australia) Fellowship [grant number APP1136250]. The authors would like to acknowledge the valuable feedback of three anonymous reviewers, which helped us to improve the quality of the original manuscript.

\bibliography{references}

\end{document}